\documentclass[runningheads]{llncs}
\usepackage[T1]{fontenc}
\usepackage{graphicx,verbatim}

\usepackage{amsmath}
\usepackage{dsfont}
\usepackage{bm}

\usepackage{multirow}
\usepackage{tabularx}
\usepackage{booktabs}

\usepackage[table, xcdraw]{xcolor}
\usepackage{colortbl}
\usepackage{hyperref}
\usepackage{glossaries}
\usepackage[ruled, vlined]{algorithm2e} 
\usepackage{color}

\begin{document}
\title{From Transthoracic to Transesophageal: Cross-Modality Generation using LoRA Diffusion}
%

\author{Emmanuel Oladokun\inst{1}\orcidID{0000-0003-2935-1552} \and Yuxuan Ou\inst{1}\orcidID{0009-0003-9406-6050} \and Anna Novikova\inst{2}\orcidID{ 0009-0004-4085-6548} \and  Daria Kulikova\inst{2}\orcidID{0000-0002-3087-5874} \and   Sarina Thomas\inst{2}\orcidID{0000-0002-1202-0856} \and  Jurica Šprem\inst{2}\orcidID{0000-0002-9165-0847} \and  Vicente Grau\inst{1}\orcidID{0000-0001-8139-3480}}  
\authorrunning{E. Oladokun et al.}
\institute{University of Oxford \and GE HealthCare, Cardiovascular Ultrasound R\&D\\ \email{emmanuel.oladokun@eng.ox.ac.uk}}

\maketitle              
\begin{abstract}

Deep diffusion models excel at realistic image synthesis but demand large training sets—an obstacle in data‐scarce domains like transesophageal echocardiography (TEE). While synthetic augmentation has boosted performance in transthoracic echo (TTE), TEE remains critically underrepresented, limiting the reach of deep learning in this high‐impact modality.

We address this gap by adapting a TTE‐trained, mask‐conditioned diffusion backbone to TEE with only a limited number of new cases and adapters as small as $10^5$ parameters. Our pipeline combines Low-Rank Adaptation with MaskR$^2$, a lightweight remapping layer that aligns novel mask formats with the pretrained model’s conditioning channels. This design lets users adapt models to new datasets with a different set of anatomical structures to the base model's original set.

Through a targeted adaptation strategy, we find that adapting only MLP layers suffices for high‐fidelity TEE synthesis. Finally, mixing less than 200 real TEE frames with our synthetic echoes improves the dice score on a multiclass segmentation task, particularly boosting performance on underrepresented right‐heart structures.   Our results demonstrate that (1) semantically controlled TEE images can be generated with low overhead, (2) MaskR$^2$ effectively transforms unseen mask formats into compatible formats without damaging downstream task performance, and (3) our method generates images that are effective for improving performance on a downstream task of multiclass segmentation.


\end{abstract}

\section{Introduction}
Echocardiography (Echo) plays a pivotal role in cardiovascular care, providing a fundamental tool to evaluate and manage cardiac diseases \cite{Potter2019TheCare}. In literature, the standard usage of 'echocardiogram' often refers to transthoracic echocardiography (TTE) specifically, which captures images from outside the subject's chest.  Transesophageal Echocardiography (TEE) involves inserting a specialised probe with an ultrasound transducer into the esophagus. This allows for clearer and more precise imaging, as the esophagus is located close to the upper chambers of the heart and the probe is not occluded by the sternum and ribs \cite{Hahn2022}. TEE is used less frequently compared to TTE due to the more complex and more invasive setup, but is especially beneficial for the diagnosis of valvular diseases and guiding minimally invasive heart surgery such as the insertion of the mitral or tricuspid clip or closure of the left atrial appendage.

Although there is much research on TTE image analysis \cite{Reynaud2023Feature-ConditionedSynthesis,Nguyen2024Training-FreeSynthesis,StojanovskiEchoSegmentation}, TEE is less researched and lacks public resources. To address data scarcity, many resort to data augmentation, which has been shown to aid in the training of rigorous models with limited data \cite{Ronneberger2015U-Net:Segmentation}. Common augmentation methods such as standard geometric transformations or contrast adjustments, have limited use in echo. Moreover, geometric transforms could easily generate physiologically impossible images. Consequently, some works have resorted to training generative models to source augmented training data \cite{Reynaud2023Feature-ConditionedSynthesis,Nguyen2024Training-FreeSynthesis,StojanovskiEchoSegmentation}. 

Generative models have significantly shaped medical image analysis and generation in recent years starting with Variational Autoencoders (VAE) \cite{Kingma2013Auto-EncodingBayes}. VAEs provide a probabilistic framework for learning latent representations but are prone to blurry results which is a significant drawback for echocardiography. VAEs were followed by Generative Adversarial Networks (GAN) \cite{Goodfellow2014GenerativeNetworks} that consist of a generator part, producing images, and a discriminator part verifying that the images look realistic. However, GANs are prone to mode collapse and training instabilities. In recent years, there has been a paradigm shift towards diffusion models which were first introduced by Sohl-Dickstein et al. \cite{Sohl-DicksteinDeepThermodynamics}. Diffusion models have emerged as a powerful class of generative models, demonstrating state-of-the-art performance in image synthesis and various data-generation tasks. These models are based on a two-step process: a forward diffusion process, where noise is gradually added to the data over multiple steps, and a reverse denoising process, where a trained model gradually removes noise to reconstruct the original data. Diffusion models are highly effective for applications such as text-to-image generation and are established as a strong choice not only in the natural, but also in the medical image domain \cite{Wolleb2022DiffusionDetection,Zhou2024CascadedTranslation} for high-fidelity image generation. 




Recent literature has attempted TTE video synthesis using several public datasets and large models. Reynaud et al. \cite{Reynaud2023Feature-ConditionedSynthesis} trained a text-to-video diffusion model to generate TTE videos with a user-specified ejection fraction.  Nguyen et al. \cite{Nguyen2024Training-FreeSynthesis} synthesised TTE videos with a training-free approach using a 3D UNet. These achievements are made possible due to the availability of high-quality, rich, public TTE datasets \cite{Ouyang2019Echonet-dynamic:Learning,Leclerc2019DeepEchocardiography,MagyarRVENet:Function}. Other research such as \cite{StojanovskiEchoSegmentation} has used diffusion models to generate key frames from TTE semantic maps. However, these approaches are limited to TTE. For TEE, \cite{Oladokun2024TransesophagealModels} generated synthetic TEE images of key frames using a CycleGAN \cite{Zhu2017UnpairedNetworks} and Contrastive Unpaired Translation method \cite{Park2020ContrastiveTranslation}.  TEE is significantly underrepresented in the literature and there are very few resources (i.e. public datasets) available to enable this to change. This paper aims to tackle this underrepresentation with the following contributions:

\begin{enumerate}
\item We present a LoRA-based method for efficient training on limited echocardiography data. This enables generation of realistic synthetic TEE images that strongly respect image conditioning, helping address data scarcity in this domain.

\item We propose a targeted adaptation strategy that reveals the functional importance of different layers within a diffusion model, identifying which layers are most critical for adapting to echo data.

\item We introduce a mask adaptation scheme that transforms new semantic maps to match the expected input of pretrained base models. Despite this transformation, our synthetic data significantly boosts downstream multiclass segmentation performance, particularly for underrepresented classes.
\end{enumerate}

\section{Methods}
\subsubsection{Diffusion Models}
 In a typical diffusion model, there are two processes: forward and reverse. The forward process gradually adds Gaussian noise to a data sample ${\bm{x_0}}$ such that at time $t$ the sample $\bm{x_t}$ has the following distribution:
\begin{equation}
    q(\bm{x_t} \mid \bm{x_0}) = \mathcal{N}\left(\bm{x_t}; \mu_{t} \bm{x_0},\, \sigma_{t}^{2} \mathbf{I}\right).
\end{equation}
where $\mu_{t}$ and $\sigma_{t}$ are the mean and variance at time $t$.
If $t$ is chosen to be large enough, the image becomes indistinguishable from random noise. For the reverse process, a denoiser is trained to iteratively remove the added noise.

In this work, we make use of the Elucidated Diffusion Model (EDM) \cite{Karras2022ElucidatingModels}. EDM presents popular variants of diffusion models—such as variance-preserving, variance-exploding, and DDIM—within a unified framework that highlights key design choices contributing to generative performance. Karras et al. \cite{Karras2022ElucidatingModels} identify two major sources of error in the reverse step: inaccurate denoising by the neural network, and the discrete solver steps that follow incorrect trajectories during sampling. To mitigate these issues, they use a second-order Heun solver for the reverse step and propose a range of conditioning strategies. Given a dataset with variance $\sigma^2_{\text{data}}$ and a noise schedule with variance $\sigma^2 = \sigma^2(t)$, EDM introduces the following preconditioning steps: scale the network input by $c_{\text{in}} = \frac{1}{\sqrt{\sigma^2_{\text{data}} + \sigma^2}}$; scale the skip connections by $c_{\text{skip}} = \frac{\sigma^2}{\sigma^2_{\text{data}} + \sigma^2}$; condition the network on noise using $c_{\text{noise}} = \frac{1}{4}\ln(\sigma)$; and scale the output by $c_{\text{out}} = \frac{\sigma \cdot \sigma_{\text{data}}}{\sqrt{\sigma^2_{\text{data}} + \sigma^2}}$. The resulting EDM loss is:
\begin{equation}
    \mathds{E}_{\sigma, x, n} \| \underbrace{F_{\theta} \big( c_{\text{in}}(\sigma) \cdot \bm{\Tilde{x}}; c_{\text{noise}}(\sigma) \big)}_{\text{network output}} - \underbrace{\frac{1}{c_{\text{out}}(\sigma)} \left( y - c_{\text{skip}}(\sigma) \cdot \Tilde{\bm{x}} \right)}_{\text{effective training target}}\|_2^2
\end{equation}
where $\bm{\Tilde{x}} = \bm{x + n}$, $\bm{n} \sim  \mathcal{N}(\bm{0}, \sigma^2\bm{I})$, and $F_{\theta}$ is the network to be trained.
They show that these reparametrisations significantly improve both training efficiency and the quality of the generated images.

\subsubsection{Low-Rank Adaptation}
Low-Rank Adaptation (LoRA) \cite{HuLORA:MODELS} is an efficient method for adapting models to new tasks. Given a pretrained model with weight matrix $W_0 \in \mathds{R}^{d \times k}$, LoRA decomposes the fine-tuning update $\triangle W$ into a product of two low-rank matrices, $BA$, where $A \in \mathds{R}^{r\times k}$ and $B \in \mathds{R}^{d \times r}$, with $r \ll \min(d,k)$. During training, the base model remains frozen and only $A$ and $B$ are updated. For an input $x$, the forward pass becomes $h = W_{0}x + \frac{\alpha}{r}BAx$, where the scaling factor $\alpha$ adjusts the influence of the adapter relative to the base model. A key advantage of LoRA adapters is that they introduce no additional inference-time latency, as they can be merged with the main model after training. We leverage LoRA to efficiently adapt the base model to a different form of echocardiography.
\subsubsection{$\text{MaskR}^2$}
\begin{figure}
    \centering
    \includegraphics[width=1\linewidth]{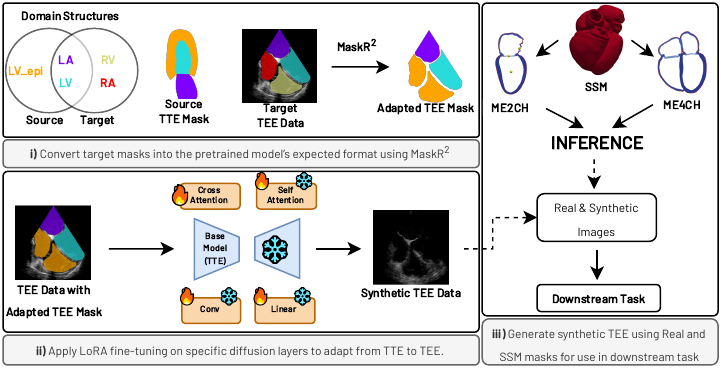}
    \caption{\textbf{TTE $\rightarrow$ TEE Pipeline.}  MaskR$^2$ is used to adapt the TEE Masks to the style expected by the TTE pretrained model. In this use case, the channel originally used for the left ventricular epicardium (LV$_{epi}$) conditioning is now responsible for generating the right-hand-side of the heart in TEE.  Next, we adapt the base model in a targeted manner to generate synthetic TEE data. After training, we perform inference using real masks and masks sampled from SSMs to generate synthetic TEE datasets. Finally, we augment existing real TEE datasets and use them for a downstream task.}
    \label{fig:tte_to_tee}
\end{figure}

One‐hot encoding (OHE) is the standard way to condition diffusion models on semantic masks: a 2D label map of size $H \times W$ becomes an $N \times H \times W$ tensor, where $N$ is the number of classes. However, OHE rigidly fixes the number of conditioning channels, forcing an advance decision on how many classes will ever be needed. If a model pretrained on a dataset with class set $X$ is to be fine-tuned on a new dataset with a different class set $Y$, either channels will be wasted  or it will not be possible to accommodate the new classes without retraining the base model. To solve this, we propose $\text{MaskR}^2$, which remaps any new-dataset labels $Y$ into the pretrained model's label space $X$ using just three simple operations - \textit{Identity, Reduce,} and \textit{Repurpose} - so that the condition architecture can remain unchanged. Concretely:
\begin{enumerate}
    \item Identity: Leave labels in $X \cap Y$ unchanged
    \item Reduce: If $|Y|>|X|$, merge extra labels in $Y \setminus X$ into 'super-classes'
    \item Repurpose: Assign (super-)classes in $Y \setminus X$ to the classes in $X \setminus Y$
    \end{enumerate}
For example, suppose a base model is trained on labels for the left atrium (LA), left ventricle (LV), and left ventricular epicardium (LV$_{epi}$), and we wish to adapt to a new dataset containing labels for LA, LV, right atrium (RA), and right ventricle (RV), MaskR$^2$ maps as follows: $\{LA \rightarrow LA, LV \rightarrow LV, \{RA, RV\} \rightarrow LV_{epi}\}$. Figure \ref{fig:tte_to_tee} part i) illustrates this mapping with real images.

\subsubsection{Data}
For training and evaluation, we utilise both an internal TEE dataset and the public CAMUS dataset \cite{Leclerc2019DeepEchocardiography} which contains TTE images. CAMUS provides 2,000 two‐ and four‐chamber echo frames at end‐systole and end‐diastole. We split these into 1400 training (70\%), 300 validation (15\%), and 300 testing (15\%) images. Our in‐house TEE collection comprises 288 image–mask pairs drawn from 71 mid‐esophageal two‐ and four‐chamber (ME2CH/ME4CH) videos. Every frame is annotated for LA and LV, and for RA and RV when visible. Two expert cardiologists with daily echocardiography experience provided the majority of labels $(\sim70\%)$, and the rest was annotated by the first author under their supervision. We allocate 196 TEE images for training, 40 for validation, and 52 for testing, ensuring an even distribution across views and cardiac phases. This disparity in dataset size reflects real‐world constraints on TEE data availability. We also sample semantic masks for the ME2CH and ME4CH views using a publicly available pipeline introduced in \cite{Oladokun2024TransesophagealModels}. This pipeline extracts planes from 3D heart statistical shape models (SSMs)  \cite{Rodero2021LinkingHeart} that correspond to standard TEE and TTE views defined by the American Society of Echocardiography \cite{Hahn2022}.
\subsubsection{Image Generation}
Figure \ref{fig:tte_to_tee} illustrates our proposed pipeline. The base model is an EDM trained at a resolution of $224 \times 224$, using a UNet with a depth of 3 \cite{PhilWangGitHubPytorch}. We augment the UNet with both self-attention and cross-attention layers. The channel dimensionality follows $64 \times [1, 2, 4]$ across the three stages. To maximise efficiency, cross-attention is applied only in layers 1 and 2, injecting conditioning information early in the generation process. To improve global structural understanding, self-attention is added at layer 3 and the bottleneck, where the receptive field is largest and features are most abstract. We apply exponential moving average (EMA) during training. The resulting UNet architecture contains $21.79$ million parameters and serves as our base model.

Firstly, we pretrain the base model on the CAMUS dataset, then freeze the model’s weights and attach LoRA adapters to facilitate efficient adaptation to the target dataset. To this end, we adopt a targeted adaptation strategy by categorising the model’s layers into five groups: \textit{Cross-Attention, Self-Attention, Convolution, Linear,} and \textit{Other}. LoRA adapters are then attached independently to each group, as well as to their combinations, to identify which subsets contribute most effectively to adaptation. For all adapters, we set $\alpha = r = 16$. Cross-attention layers are always trained, as they control how the model integrates conditioning signals—updating these weights is essential for learning from novel mask inputs. Following a hyperparameter search, all adapted models are trained identically: for 100,000 steps, with an initial learning rate of $1 \times 10^{-3}$ and cosine decay, and a batch size of 4. As a baseline, we also train a model with the same architecture as the base model from scratch on the TEE dataset.
\subsubsection{Evaluation}
Using the test set, we sampled images from all model configurations and evaluated them using common image quality metrics: FID \cite{Heusel2017GANsEquilibrium}, LPIPS \cite{Zhang2018TheMetric}, SSIM \cite{Wang2004ImageSimilarity}. Notably, FID and LPIPS are tailored for evaluating natural scenes rather than medical echos—where features like speckle noise and subtle texture matter. However, they are commonly used and reported in similar literature so we report them here for completion with the above caveats. To assess downstream utility, we augmented the real TEE training set with synthetic images generated from both real masks and publicly available SSM masks in a 1:1 ratio, then trained nnUNet \cite{Isensee2021NnU-Net:Segmentation} on each augmented set. Notably, MaskR$^2$ only needs to be applied to the generative model inputs, therefore all original classes are available for segmentation. We also trained a baseline nnUNet on real images alone. All models were evaluated on the same held-out validation and test splits of real data. We report three metrics: Global Dice, which pools true positives, false positives, and false negatives across every class and image; Class-Weighted Dice, which weights each class’s Dice by its ground-truth pixel count in the aggregate; and Per-class Dice, the separate Global Dice computed for each class independently.


\section{Results \& Discussion}
\begin{figure}[htb!]
    \centering
    \includegraphics[width=\linewidth]{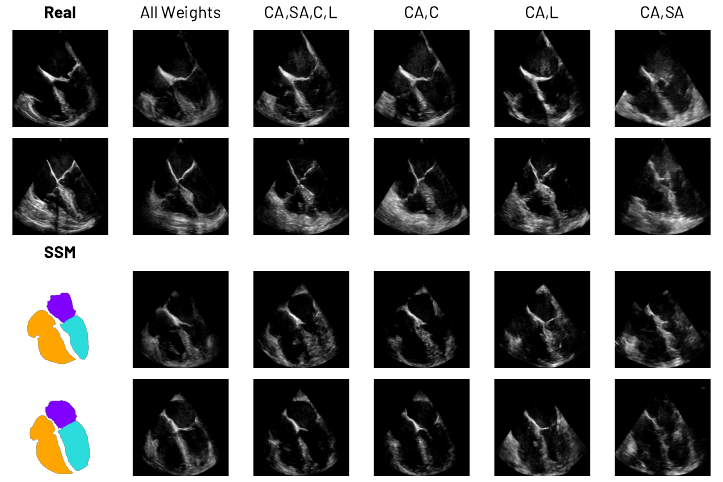}
    \caption{\textbf{Synthetic TEE Images.} This figure shows how the adapted models perform at inference using real masks from the test set and SSM masks. In all cases, the generative model sees the right side of the heart as one structure due to MaskR$^2$. 'All Weights' represents a generative model trained on TEE images from scratch. The layer groups are Cross-attention (CA), Self-attention (SA), Convolution (C), and Linear (L).}
    \label{fig:synth_comparison}
\end{figure}

Table \ref{tab:results} compares our adapter configurations on both image‐quality metrics and downstream segmentation performance under real‐mask and SSM‐mask conditioning. We first note that image quality shows a weak correlation to augmentation impact suggesting the two are not tightly coupled i.e. better looking images, according to these metrics, do not translate to more usefulness on a segmentation task. Furthermore, reducing the trainable parameters has a weak effect on image quality when we compare the adapted models to the 'All-Weights' model. This demonstrates that the adapters are able to leverage the base model's prior knowledge from TTE data to learn to generate TEE datasets with very few parameters. Next, we note that all adapters generalise well to out‐of‐distribution SSM masks: FID increases only marginally under SSM conditioning—and for the \{CA, SA\} adapter it remains unchanged. Such small degradations are encouraging, especially since a perfect FID is unattainable when comparing across different distributions. In Figure \ref{fig:synth_comparison}, we compare synthetic echoes generated by our adapters using in-distribution real masks from the held-out test set and out-of-distribution SSM masks that the model never saw during training. Despite using only up to 11\% of the original parameters, the adapters produce images with high visual fidelity, and the SSM-conditioned outputs remain anatomically plausible despite the domain gap. We observe that all models except \{CA, SA\} are capable of generating the right side of the heart well including valves. The valves are less pronounced in the SSM generated images as there are no gaps between the RA and RV. 
\begin{table}[htb]
\caption{\textbf{Generation \& Multiclass Segmentation Results.} This table summarises the performance of our LoRA‐adapted models on both image‐quality metrics and a downstream multiclass segmentation task. “All Weights” refers to the model purely trained on TEE. Each adapter updates only the specified layer groups—Cross‐attention (CA), Self‐attention (SA), Convolution (Conv), and Linear—via LoRA. “Mask Source” indicates whether the synthetic echoes were generated with real TEE masks or SSM masks. The “Per-class Dice” column reports the Dice score for each chamber, shown here as the delta relative to the baseline model. \textbf{Bold} highlights the overall best scores, and colored entries mark the top performer for each individual class.\textsuperscript{\dag}Best FID Score in \cite{Oladokun2024TransesophagealModels}. \textsuperscript{\ddag}Scores achieved from nnUNet trained on purely real data.}
\label{tab:results}
\resizebox{\columnwidth}{!}{%
\begin{tabular}{@{}lllcccccc@{}}
\toprule
\multicolumn{3}{c}{\textbf{Generative Model}} &
  \multicolumn{3}{c}{\textbf{Image Quality Metrics}} &
  \multicolumn{3}{c}{\textbf{Segmentation Scores}} \\ \midrule
\begin{tabular}[c]{@{}l@{}}Trainable\\ Groups \end{tabular} &
  \begin{tabular}[c]{@{}l@{}}Trainable\\ Params (M/\%)\end{tabular} &
  \multicolumn{1}{l|}{\begin{tabular}[c]{@{}l@{}}Mask\\ Source\end{tabular}} &
  FID $(\downarrow)$ &
  LPIPS $(\downarrow)$ &
  \multicolumn{1}{c|}{SSIM $(\uparrow)$} &
  \begin{tabular}[c]{@{}c@{}}Global\\ Dice $(\uparrow)$\end{tabular} &
  \begin{tabular}[c]{@{}c@{}}Class-weighted\\ Dice $(\uparrow)$\end{tabular} &
  \begin{tabular}[c]{@{}c@{}}Per-class Dice $(\uparrow)$\\ \{LA, LV, RV, RA\}\end{tabular} \\ \midrule
 &
   &
  \multicolumn{1}{l|}{\cellcolor[HTML]{EFEFEF}Real} &
  \cellcolor[HTML]{EFEFEF}155.7 &
  \cellcolor[HTML]{EFEFEF}0.31 &
  \multicolumn{1}{c|}{\cellcolor[HTML]{EFEFEF}0.55} &
  \cellcolor[HTML]{EFEFEF}89.05 &
  \cellcolor[HTML]{EFEFEF}88.78 &
  \cellcolor[HTML]{EFEFEF}\{+0.22, +1.75, +5.82, +2.97\} \\
\multirow{-2}{*}{All Weights} &
  \multirow{-2}{*}{21.79/100\%} &
  \multicolumn{1}{l|}{SSM} &
  176.5 &
  - &
  \multicolumn{1}{c|}{-} &
  88.61 &
  89.41 &
  \{+0.63, +1.46, +3.87, +2.56\} \\ \midrule
 &
   &
  \multicolumn{1}{l|}{\cellcolor[HTML]{EFEFEF}Real} &
  \cellcolor[HTML]{EFEFEF}117.7 &
  \cellcolor[HTML]{EFEFEF}0.31 &
  \multicolumn{1}{c|}{\cellcolor[HTML]{EFEFEF}0.54} &
  \cellcolor[HTML]{EFEFEF}88.53 &
  \cellcolor[HTML]{EFEFEF}88.79 &
  \cellcolor[HTML]{EFEFEF}\{+0.07, +1.15, +4.48, \textcolor[HTML]{FF0000}{\textbf{+3.49}}\} \\
\multirow{-2}{*}{\begin{tabular}[c]{@{}l@{}}\{CA, Linear\\ SA, Conv\}\end{tabular}} &
  \multirow{-2}{*}{2.70/11\%} &
  \multicolumn{1}{l|}{SSM} &
  120.3 &
  - &
  \multicolumn{1}{c|}{-} &
  88.21 &
  89.06 &
  \{+0.97, +0.56, +3.53, +2.91\} \\ \midrule
 &
   &
  \multicolumn{1}{l|}{\cellcolor[HTML]{EFEFEF}Real} &
  \cellcolor[HTML]{EFEFEF}134.5 &
  \cellcolor[HTML]{EFEFEF}0.31 &
  \multicolumn{1}{c|}{\cellcolor[HTML]{EFEFEF}0.55} &
  \cellcolor[HTML]{EFEFEF}88.14 &
  \cellcolor[HTML]{EFEFEF}88.75 &
  \cellcolor[HTML]{EFEFEF}\{-0.11, +1.05, +4.23, -0.14\} \\
\multirow{-2}{*}{\{CA, Conv\}} &
  \multirow{-2}{*}{2.13/9\%} &
  \multicolumn{1}{l|}{SSM} &
  160 &
  - &
  \multicolumn{1}{c|}{-} &
  88.28 &
  89.77 &
  {\color[HTML]{000000} \{\textcolor[HTML]{7F00FF}{\textbf{+1.03}}, +1.15, +3.08, +0.68\}} \\ \midrule
 &
   &
  \multicolumn{1}{l|}{\cellcolor[HTML]{EFEFEF}Real} &
  \cellcolor[HTML]{EFEFEF}154.4 &
  \cellcolor[HTML]{EFEFEF}0.33 &
  \multicolumn{1}{c|}{\cellcolor[HTML]{EFEFEF}0.53} &
  \cellcolor[HTML]{EFEFEF}{\color[HTML]{000000} 89.40} &
  \cellcolor[HTML]{EFEFEF}89.72 &
  \cellcolor[HTML]{EFEFEF}\{-0.11, \textcolor[HTML]{2ADCDC}{\textbf{+3.43}}, +3.64, +0.69\} \\
\multirow{-2}{*}{\{CA, Linear\}} &
  \multirow{-2}{*}{0.69/3\%} &
  \multicolumn{1}{l|}{SSM} &
  164.3 &
  - &
  \multicolumn{1}{c|}{-} &
  \textbf{89.60} &
  \textbf{90.14} &
  \{+0.92, +2.62, \textcolor[HTML]{D4DC7F}{\textbf{+5.90}}, +3.09\} \\ \midrule
 &
   &
  \multicolumn{1}{l|}{\cellcolor[HTML]{EFEFEF}Real} &
  \cellcolor[HTML]{EFEFEF}152.3 &
  \cellcolor[HTML]{EFEFEF}0.35 &
  \multicolumn{1}{c|}{\cellcolor[HTML]{EFEFEF}0.48} &
  \cellcolor[HTML]{EFEFEF}89.38 &
  \cellcolor[HTML]{EFEFEF}89.51 &
  \cellcolor[HTML]{EFEFEF}\{+0.56, +3.24, +2.70, +0.28\} \\
\multirow{-2}{*}{\{CA, SA\}} &
  \multirow{-2}{*}{0.51/2\%} &
  \multicolumn{1}{l|}{SSM} &
  151.4 &
  - &
  \multicolumn{1}{c|}{-} &
  87.56 &
  88.45 &
  \{+0.83, +0.15,+1.90, -0.57\} \\ \midrule
\textbf{Baselines} &
  - &
  - &
  188\textsuperscript{\dag} &
  - &
  - &
  87.16\textsuperscript{\ddag} &
  88.00\textsuperscript{\ddag} &
  \{94.78, 86.65, 70.83, 84.71\} \\ \bottomrule
\end{tabular}%
}
\end{table}

On the downstream task, all augmented datasets outperform the baseline trained solely on real images, confirming that segmentation benefits from our synthetic images. The \{CA, Linear\} configuration shows the largest improvement in overall performance as well as the best performance when we compare each mask source independently. The 'Linear' group mainly consists of MLP layers which suggests that these layers are the most important for learning features that are most useful for segmentation. For all but one segmentation model, the class-weighted dice exceeds the global dice, indicating that these models perform better on the more prevalent classes. When we examine the per‐class Dice—which computes a separate Global Dice for each chamber—the underrepresented right‐heart structures consistently gain more from synthetic augmentation than the left. For example, the right ventricle sees improvements ranging from 1.9 to 5.9 Dice points. Crucially, these gains occur even though MaskR$^2$ collapses RA and RV into a single super‐class, demonstrating that the adapted models can generate synthetic images capable of enhancing right‐side segmentation without explicitly distinguishing those two chambers.

Overall, our results indicate that LoRA adapters are able to harness information learned from the more prevalent TTE and build upon it to generate useful TEE images with as little as around 510,000 parameters. Furthermore, MaskR$^2$ is able to effectively map new mask formats into the base model's expected conditioning space without compromising downstream performance.

\section{Conclusion}
We propose a lightweight, data-efficient pipeline that adapts a TTE-trained diffusion model to TEE via LoRA using minimal TEE data. By conditioning on semantic masks—and using MaskR$^2$ to remap novel mask formats into the model’s original channel space—we achieve fine-grained control over anatomy even when new structures or mask conventions arise. We show that our synthetic TEE images are both perceptually realistic and structurally faithful: when used to augment real TEE cases, they boost multiclass segmentation Dice score, with the greatest gains on underrepresented right-heart chambers. In doing so, we validate the practical use of pretrained diffusion models for specialised echo imaging. Moreover, because the SSM masks we employ are publicly available, our approach can be readily adopted by others. Finally, this adaptable framework is modality‐agnostic and can be applied to other imaging domains wherever mask‐conditioned synthesis is desired.

\section{Acknowledgements}
The authors would like to acknowledge the use of the University of Oxford
Advanced Research Computing (ARC) facility (http://dx.doi.org/10.5281/zenodo.22558)
and funding from the EPSRC CDT in Sustainable Approaches to Biomedical Science: Responsible and Reproducible Research (SABS:R$^{3}$)

%
%
%
\clearpage
\bibliographystyle{splncs04}
\bibliography{miccai2025_ref}
\end{document}